\documentclass[twocolumn,times]{aastex63} 

\usepackage{xcolor}
\definecolor{green}{HTML}{33CC33}
\definecolor{red}{HTML}{FF3300}
\definecolor{blue}{HTML}{3333FF}

\hypersetup{colorlinks=true,citecolor=blue,urlcolor=blue}

\received{Juli 26, 2019}
\revised{August 23, 2019}
\accepted{October 28, 2019}
\submitjournal{AJ}

\usepackage{amssymb}
\usepackage{amsmath}
\usepackage{xspace}
\usepackage{epstopdf}
\usepackage{url}
\usepackage{array,multirow}
\usepackage{enumitem}
\usepackage[english]{babel}

\usepackage{mathrsfs}
\usepackage{graphicx}
\usepackage{listings}

\usepackage[varg]{txfonts}
\usepackage[normalem]{ulem}

\usepackage{blindtext}
\usepackage{lipsum} 

\usepackage{subfigure}
\usepackage[T1]{fontenc}
\usepackage{ae,aecompl}

\usepackage{threeparttable}
\usepackage{booktabs}

\newcommand{\ie}{i.e.\@\xspace} 
\newcommand{\eg}{e.g.\@\xspace} 

\renewcommand{\arraystretch}{1.}

\renewcommand{\eqref}[1]{Equation~\ref{#1}}
\newcommand{\fref}[1]{Figure~\ref{#1}}
\newcommand{\tref}[1]{Table~\ref{#1}}
\newcommand{\sref}[1]{Section~\ref{#1}}

\usepackage{natbib,twoopt}

\newcommand{\numax}{\ensuremath{\nu_{\rm max}}\xspace}
\newcommand{\dnu}{\ensuremath{\Delta\nu}\xspace}
\newcommand{\kp}{\emph{Kepler}\xspace}

\newcommand{\teff}{\ensuremath{T_{\rm eff}}\xspace}

\newcommand{\logg}{\ensuremath{\log g}\xspace}
\newcommand{\feh}{\ensuremath{\rm [Fe/H]}\xspace}

\numberwithin{equation}{section}

\defcitealias{K293}{V16a} 

\makeatletter
\def\maketag@@@#1{\hbox{\m@th\normalfont\normalsize#1}}
\makeatother

\bibpunct{(}{)}{;}{a}{}{,} 

\makeatletter
\newcommand*\mysize{%
  \@setfontsize\mysize{5.7}{8.0}%
}
\makeatother

\makeatletter
\newcommand*\tabsize{%
  \@setfontsize\tabsize{7.}{8.0}%
}
\makeatother

\makeatletter
\newcommand\footnoteref[1]{\protected@xdef\@thefnmark{\ref{#1}}\@footnotemark}
\makeatother

\shorttitle{The K2-93 Multiplanet System}
\shortauthors{Lund et al.}

\begin{document}

\title{Asteroseismology of the Multiplanet System K2-93}

\correspondingauthor{Mikkel N. Lund}
\email{mikkelnl@phys.au.dk}

\author[0000-0001-9214-5642]{Mikkel~N.~Lund}
\affiliation{Stellar Astrophysics Centre, Department of Physics and Astronomy, Aarhus University, Ny Munkegade 120, DK-8000 Aarhus C, Denmark}

\author[0000-0001-7880-594X]{Emil~Knudstrup}
\affiliation{Stellar Astrophysics Centre, Department of Physics and Astronomy, Aarhus University, Ny Munkegade 120, DK-8000 Aarhus C, Denmark}

\author[0000-0002-6137-903X]{V{\'{\i}}ctor~Silva~Aguirre}
\affiliation{Stellar Astrophysics Centre, Department of Physics and Astronomy, Aarhus University, Ny Munkegade 120, DK-8000 Aarhus C, Denmark}

\author[0000-0002-6163-3472]{Sarbani~Basu}
\affiliation{Department of Astronomy, Yale University, PO Box 208101, New Haven, CT 06520-8101, USA}

\author[0000-0003-1125-2564]{Ashley Chontos}
\altaffiliation{NSF Graduate Research Fellow}
\affiliation{Institute for Astronomy, University of Hawai`i, 2680 Woodlawn Drive, Honolulu, HI 96822, USA}

\author[0000-0002-6956-1725]{Carolina~Von~Essen}
\affiliation{Stellar Astrophysics Centre, Department of Physics and Astronomy, Aarhus University, Ny Munkegade 120, DK-8000 Aarhus C, Denmark}
\affiliation{Astronomical Observatory, Institute of Theoretical Physics and Astronomy, Vilnius University, Sauletekio av. 3, 10257, Vilnius, Lithuania}

\author[0000-0002-5714-8618]{William~J.~Chaplin}
\affiliation{School of Physics and Astronomy, University of Birmingham, Edgbaston, Birmingham, B15 2TT, UK}
\affiliation{Stellar Astrophysics Centre, Department of Physics and Astronomy, Aarhus University, Ny Munkegade 120, DK-8000 Aarhus C, Denmark}

\author[0000-0001-6637-5401]{Allyson~Bieryla}
\affiliation{Center for Astrophysics | Harvard-Smithsonian, 60 Garden Street Cambridge, MA 02138, USA}

\author[0000-0003-2688-7511]{Luca~Casagrande}
\affiliation{Research School of Astronomy and Astrophysics, Mount Stromlo Observatory, The Australian National University, ACT 2611, Australia}

\author[0000-0001-7246-5438]{Andrew~Vanderburg}
\altaffiliation{NASA Sagan Fellow}
\affiliation{Department of Astronomy, The University of Texas at Austin, Austin, TX 78712, USA}

\author[0000-0001-8832-4488]{Daniel Huber}
\affiliation{Institute for Astronomy, University of Hawai`i, 2680 Woodlawn Drive, Honolulu, HI 96822, USA}

\author[0000-0002-7084-0529]{Stephen~R.~Kane}
\affiliation{Department of Earth and Planetary Sciences, University of California, Riverside, CA 92521, USA}

\author[0000-0003-1762-8235]{Simon~Albrecht}
\affiliation{Stellar Astrophysics Centre, Department of Physics and Astronomy, Aarhus University, Ny Munkegade 120, DK-8000 Aarhus C, Denmark}

\author[0000-0001-9911-7388]{David~W.~Latham}
\affiliation{Center for Astrophysics | Harvard-Smithsonian, 60 Garden Street Cambridge, MA 02138 USA}

\author[0000-0002-4290-7351]{Guy~R.~Davies}
\affiliation{School of Physics and Astronomy, University of Birmingham, Edgbaston, Birmingham, B15 2TT, UK}
\affiliation{Stellar Astrophysics Centre, Department of Physics and Astronomy, Aarhus University, Ny Munkegade 120, DK-8000 Aarhus C, Denmark}

\author[0000-0002-7733-4522]{Juliette~C.~Becker}
\altaffiliation{NSF Graduate Research Fellow}
\affiliation{Astronomy Department, University of Michigan, 1085 S University Avenue, Ann Arbor, MI 48109, USA}

\author[0000-0001-8812-0565]{Joseph~E.~Rodriguez}
\affiliation{Center for Astrophysics | Harvard-Smithsonian, 60 Garden Street Cambridge, MA 02138 USA}

\begin{abstract}
We revisit the analysis of the bright multiplanet system K2-93, discovered with data taken by the K2 mission. This system contains five identified planets ranging in size from sub-Neptune to Jupiter size. The K2 data available at the discovery of the system only showed single transits for the three outer planets, which allowed weak constraints to be put on their periods. As these planets are interesting candidates for future atmospheric studies, a better characterization of the host star and tighter constraints on their orbital periods are essential.
Using new data from the K2 mission taken after the discovery of the system, we perform an asteroseismic characterization of the host star. We are able to place strong constraints on the stellar parameters and obtain a value for the stellar mass of $1.22^{+0.03}_{-0.02}\, \rm M_{\odot}$, a stellar radius of $1.30\pm 0.01\, \rm R_{\odot}$, and an age of $2.07^{+0.36}_{-0.27}$ Gyr. Put together with the additional transits identified for two of the three outer planets, we constrain the orbital periods of the outer planets and provide updated estimates for the stellar reflex velocities induced by the planets.  
\end{abstract}


\section{Introduction} \label{sec:intro}
The K2-93 system was first discovered by \citet{K293} (hereafter \citetalias{K293}) from data obtained during Campaign 5 (C5) of the K2 mission \citep{K2Howell}. This analysis revealed a system with five transiting planets, two inner sub-Neptune-sized planets, and three outer planets ranging from Neptune to Jupiter size. The three outer planets only showed a single transit in the C5 data, hence their periods could only be loosely predicted based on the available stellar parameters and dynamical stability considerations.
Based on K2 data from Campaign 18 (C18), \citet{bernado} and \citet{becker} discovered additional transits for two (``d'' and ``f'') of the three outer planets, which allowed stronger constraints to be placed on their periods.

The system is particularly interesting because the stellar host, HIP 41378 (EPIC 211311380), is relatively bright, with a \textit{V}-band magnitude of $8.93$ (and \textit{JHK$_s$} magnitudes from $7.7-8.0$), enabling follow-up studies from ground. In contrast, the only other multi-planet transiting systems with periods beyond that of the outermost planet ``f'', of which there are four from the \kp mission\footnote{found from searching the NASA exoplanet archive (\url{https://exoplanetarchive.ipac.caltech.edu/index.html}) on 12 April 2019.}, all have \textit{JHK$_s$} magnitudes of the order ${\sim}12$. Only one of these is also a confirmed multi-planet system. The fact that the star is bright and that the Jupiter-sized ``f'' planet orbits far from its host means that in addition to causing a deep transit it is an ideal target for transit transmission spectroscopy. As discussed by \citetalias{K293} it may even be possible to measure the planetary oblateness, because the planetary orbit will not have synchronized with the orbital period.

In this paper we present the detection of solar-like oscillations in K2-93 using the C18 short-cadence data. Using asteroseismology  \citep{2010aste.book.....A} we provide a significant improvement in the characterization of the host star of this benchmark system, including the planet radii and orbital periods for planets ``d'' and ``f''. We also use data from TESS to provide additional constraints on the period of planet ``e''.

\section{Data} \label{sec:data}
HIP 41378 was first observed by the K2 mission during C5 in long-cadence mode (LC; $\Delta t \sim 30\,\rm min$). The star was observed again in LC in C18, and was also observed in short-cadence (SC; $\Delta t \sim 1\,\rm min$) mode\footnote{\url{https://keplerscience.arc.nasa.gov/k2-approved-programs.html\#campaign-18}} to search for asteroseismic signals.


Light curves were constructed from pixel-data downloaded from the KASOC database\footnote{\url{www.kasoc.phys.au.dk}}, extracted using the \textit{K2P$^2$} pipeline \citep{k2p2} and corrected using the KASOC filter \citep[][]{2014MNRAS.445.2698H}, which iteratively corrects for both known planetary transits, long-term trends, sharp features, and the characteristic ${\sim}6$-hour systematic of the K2 mission \citep{2014PASP..126..948V,2016van.cleve.pasp}. 

\begin{figure*}
\includegraphics[width=\textwidth]{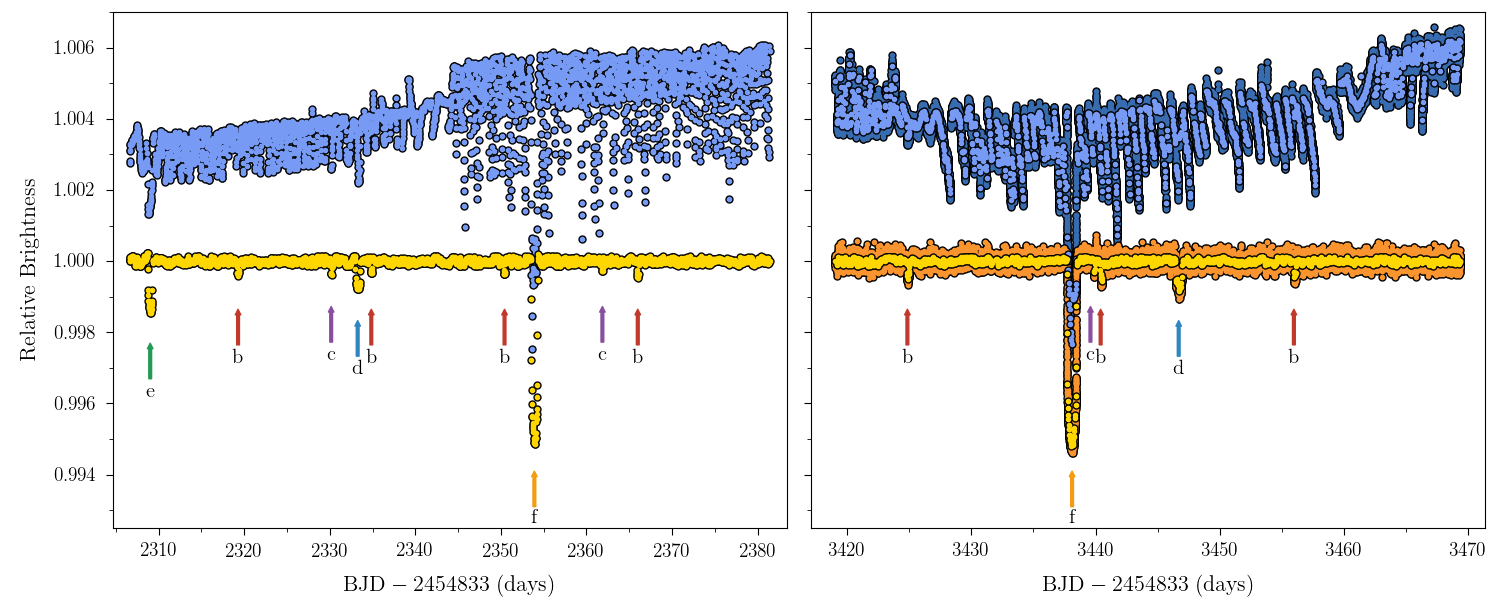}
\caption{Light curve for HIP 41378 obtained during C5 (left) and C18 (right). The blue points show the raw photometry (offset by $0.04$), while the yellow points show the light curve after correcting for the K2 systematics. Light colored points show data in LC, dark points (only in C18) show the SC data. Each of the identified transits of the five planets have been indicated with arrows.}
\label{fig:lc}
\end{figure*}

\fref{fig:lc} shows the raw and corrected light curve for HIP 41378, though without correcting for the planetary transits as done for the seismic analysis. As seen, the outer planets ``d'' and ``f'' transit again in C18, where only a single transit was available before from C5. Planet ``e'' unfortunately does not transit again. Our planetary analysis is described in \sref{sec:planet}.

\section{Analysis} \label{sec:ana}

\subsection{Asteroseismic parameters}\label{sec:aa}

\begin{figure*}
\includegraphics[width=\textwidth]{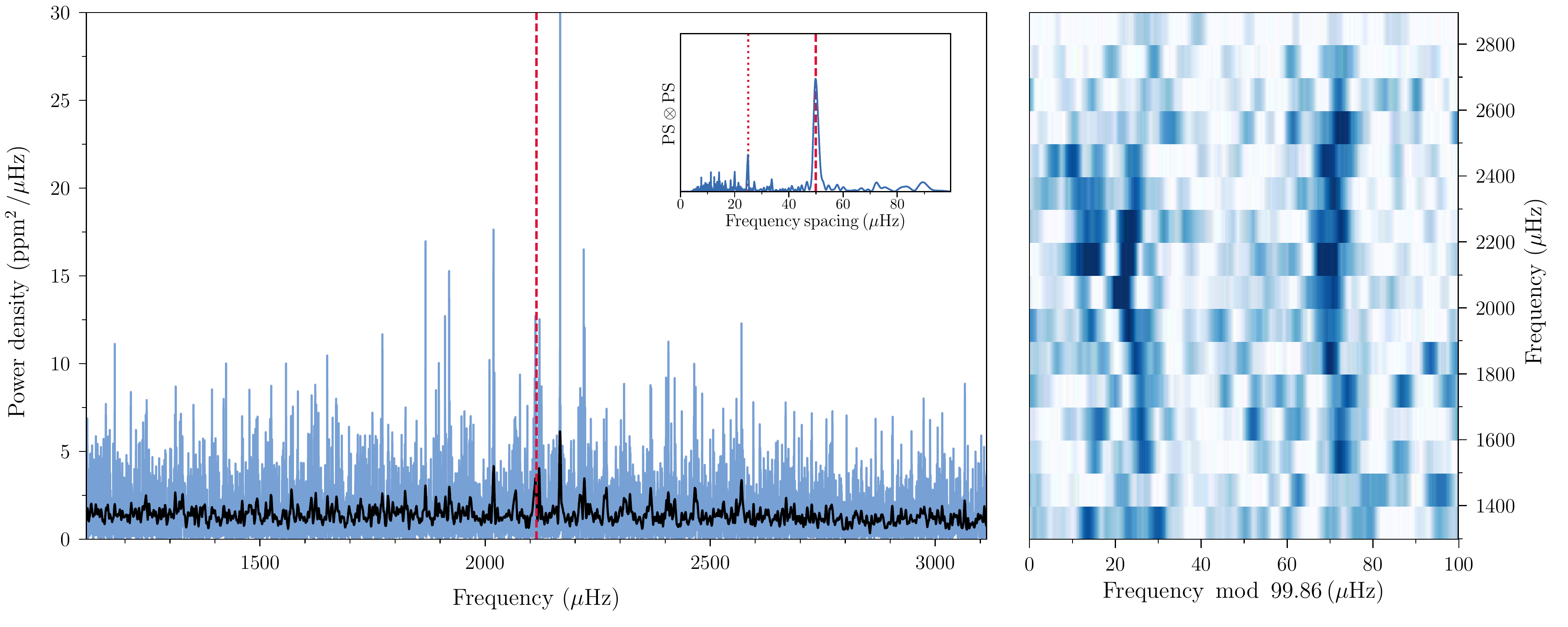}
\caption{Power density spectrum (PDS; left) and \'{e}chelle diagram (right) of HIP 41378. The black line shows a smoothed version of the PDS and the red dashed line indicates the value of \numax. The insert in the PDS shows the $\rm {PS\otimes PS}$ of a region centered on the measured \numax; the peak in the $\rm {PS\otimes PS}$ marked by a dashed red line gives the identified value for $\dnu/2$, while the dotted line shows the value of $\dnu/4$.}
\label{fig:ps}
\end{figure*}

We determined values for the average asteroseismic parameters \dnu, the average spacing in frequency between consecutive modes of the same angular degree, and \numax, the frequency of maximum mode power. The value of \numax was determined following the procedure of \citet{keystone} using a fit to the stellar granulation background including a Gaussian power hump to account for the excess power from oscillations; we determine a value of $\rm \numax=2114\pm 38\,\, \mu Hz$. \dnu was determined from the $\dnu/2$ peak of the power-of-power spectrum centered on \numax, after first having corrected for the stellar granulation background; we determine a value of $\rm \dnu=99.86\pm 2.48\,\,\mu Hz$ (see \fref{fig:ps}).

As seen from \fref{fig:ps}, especially clear from the \'{e}chelle diagram in the right panel \citep{2011arXiv1107.1723B}, individual modes of oscillation are readily visible for this star. We extracted information on the individual modes using the peak-bagging procedure outlined in \citet{legacy} (see also \citet{DaviesKAGES}). The mode identification was done by visual inspection of the power density spectrum (PDS), but we note that the obtained value $\epsilon\approx1.29$ (observationally $\dnu(\epsilon -1)$ gives the horizontal position of the $l=0$ ridge in the \'{e}chelle diagram) from our preferred identification matches predictions from \citet{2012ApJ...751L..36W} based on the stellar \teff.

\begin{figure}
\centering
\includegraphics[width=\columnwidth]{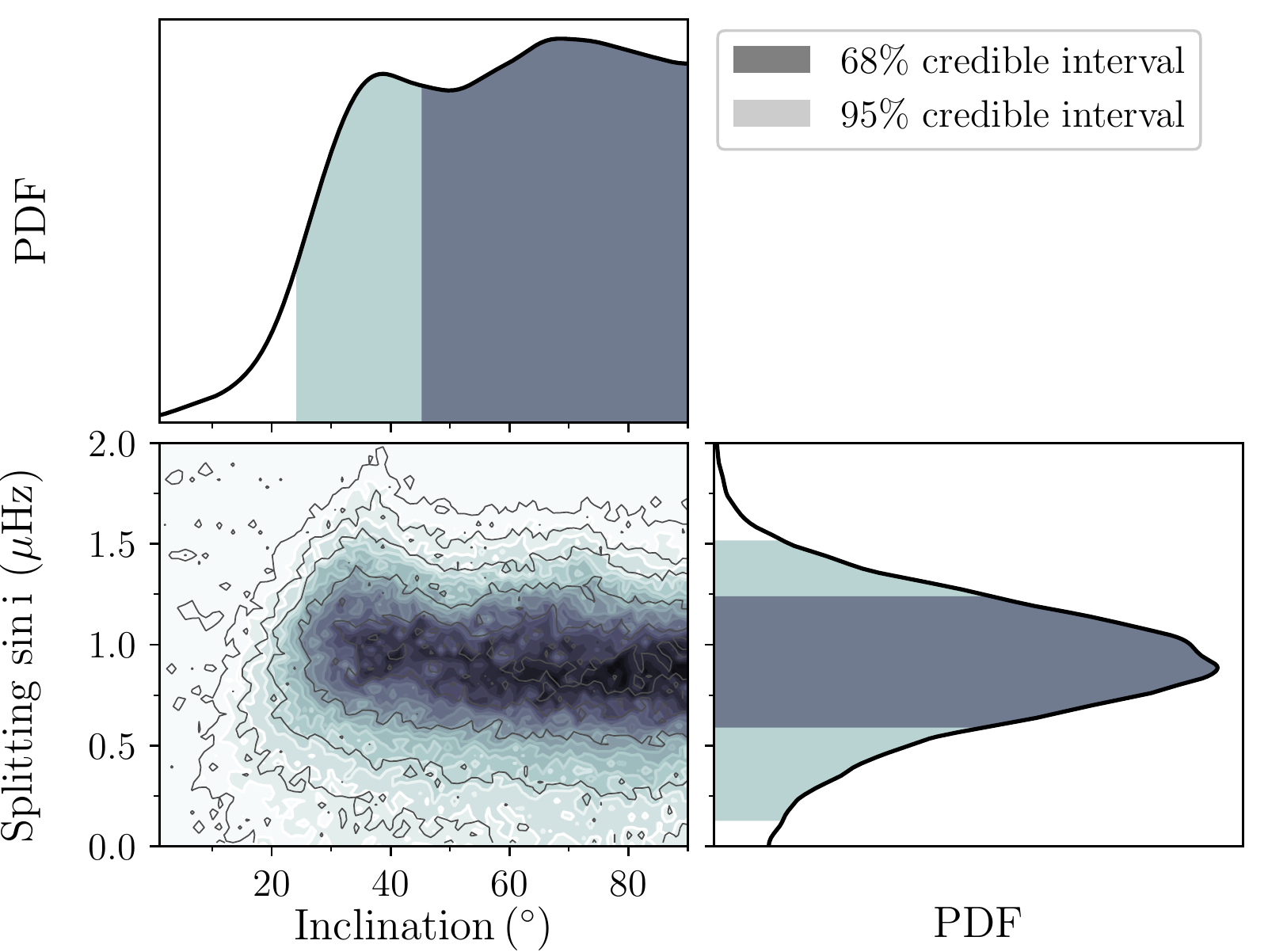}
\caption{Correlation map of the stellar inclination and projected splitting from the peak-bagging analysis. }
\label{fig:inc}
\end{figure}

As part of the peak-bagging, values are determined for the stellar inclination and the mode splitting from the stellar rotation \citep{2014ApJ...782...14V}, the former of which is particularly important to assess the obliquity of the planetary system \citep{2014A&A...570A..54L,2016ApJ...819...85C}. We fitted these parameters in projected splitting and in $\cos i_{\star}$ on which we adopted a flat prior consistent with an isotropic distribution. \fref{fig:inc} shows the correlation map between the projected splitting and the stellar inclination, here a horizontal line corresponds to a specific value for the projected rotational velocity $v\sin i_{\star}$ when taking into account the stellar radius. We determine a posterior median value for the projected splitting of ${\rm \nu_s} \sin i_{\star} = \rm 0.90\pm 0.32\,\, \mu Hz$ (with the splitting $\rm \nu_s$ given by the inverse of the stellar rotation period), and an inclination of $i_{\star}\geqslant 45^{\circ}$ as the lower limit of the $68\%$ highest probability density interval. The projected rotational velocity $v\sin i$ found from combining the fitted projected splitting with the asteroseismic radius has a value of $5.1\pm 1.8\,\, \rm km s^{-1}$ -- this is consistent with the spectroscopic value from the Stellar Parameters Classification tool \citep[SPC; see][]{2012Natur.486..375B}, especially considering that the SPC value for $v\sin i_{\star}$ will contain a contribution from macroturbulence. Assuming a contribution from macroturbulence of ${\sim}5\,\, \rm km s^{-1}$ \citep{2014MNRAS.444.3592D}, and subtracting this in quadrature from the reported SPC value, results in a $v\sin i_{\star}$ from rotation of ${\sim}5.1\,\, \rm km s^{-1}$, in full agreement with the seismic value. Only a weak constraint can be placed on the stellar inclination, which is inconsistent with a highly misaligned system, and a projected obliquity, \eg, from Rossiter-McLaughlin measurements, is required to fully constrain the system geometry \citep{2005ApJ...631.1215W,2013ApJ...771...11A}.

\subsection{Spectroscopy and Infrared Flux Method (IRFM)}\label{sec:spec}
We rederived spectroscopic parameters using spectra from the 1.5 m Tillinghast telescope at the F.~L.~Whipple Observatory, which were obtained for the original analysis by \citetalias{K293}. The spectra from TRES were analyzed using the SPC where the value of \logg was iterated based on \numax \citep[see][]{1991ApJ...368..599B,2014ApJ...783..123C} to decrease the impact on uncertainties from correlations between \teff, \logg, and \feh. Following \citet[][]{2012ApJ...757..161T} we add systematic uncertainties of $\pm 59\, \rm K$ and $\pm 0.062 \, \rm dex$ in quadrature to the derived uncertainties on \teff and \feh. The updated spectroscopic parameters are listed in \tref{tab:spec}. The main change is seen for \teff which with the seismic \logg is $91\,$K higher than the solution of \citetalias{K293}.

\begin{deluxetable}{lr}
\tablecaption{Spectroscopic and IRFM parameters.\label{tab:spec}}
\tablehead{\colhead{Parameter} & \colhead{Value}  } 
\startdata
Effective temperature from SPC, \teff (K) & $6290\pm 77$\\
Metallicity from SPC, \feh (dex) & $-0.05\pm 0.10$\\
Projected rotation speed,$^{\dagger}$ $v\sin i$ ($\rm km s^{-1}$) & $7.1\pm 0.5$\\
Surface gravity from SPC, $\log g$ (cgs) & $4.29\pm 0.10$\\
Effective temperature IRFM, \teff & $6347\pm 70$\\
Angular diameter, $\theta$ (mas) & $0.114\pm0.002$\\
\enddata
\tablecomments{\footnotesize the \logg values from SPC and the IRFM are obtained using the \teff values combined with \numax. $^{\dagger}$The $v\sin i$ from SPC also contain a broadening contribution from macroturbulence, likely resulting in a $v\sin i$ of the order ${\sim}5.1\,\rm km s^{-1}$.}
\end{deluxetable}

As a sanity check of the spectroscopic \teff we also determined this using the IRFM \citep[][]{2014MNRAS.439.2060C}. We follow the procedure outlined in \citet{keystone}, and fix in the determination \feh to the spectroscopic value and adopt an interstellar reddening of zero. Uncertainties in the reddening and metallicity are propagated to the IRFM parameters using a Monte Carlo analysis. The derived parameters are listed in \tref{tab:spec} -- as seen, the agreement between the two determinations of \teff is excellent and for both estimates within their $1-\sigma$ uncertainty.

\subsection{Asteroseismic modeling}\label{sec:am}
Before proceeding with the modeling of the extracted individual frequencies it is important to consider Doppler shifts of the frequencies from the radial velocity (RV) of the star \citep{2014MNRAS.445L..94D}. Based on \textit{Gaia} DR2 \citep{gaia2018}, the star has an RV of $\rm 50.42\pm 0.37\,\,km\, s^{-1}$, which for the range of oscillation frequencies observed would result in Doppler shifts from $\rm 0.26$ to $0.45\,\,\mu Hz$. In our case this is below the general uncertainty on the frequencies, but as the shift is systematic we account for it and thereby ensure that the uncertainty on the RV measurement is propagated to the adopted frequency values. We note that the shifts had a negligible effect in our case, and will in general have minimal importance when modeling frequency difference ratios rather than the frequencies themselves.

We model the star using The BAyesian STellar Algorithm \citep[BASTA;][]{2015MNRAS.452.2127S,2017ApJ...835..173S} with evolution models computed with the Garching Stellar Evolution Code \citep[GARSTEC;][]{2008Ap&SS.316...99W} and frequencies computed with the Aarhus adiabatic oscillation package \citep[ADIPLS;][]{2008Ap&SS.316..113C}. With BASTA, modeling was run using both frequency separation ratios $r_{010}$ and $r_{02}$ \citep{2003A&A...411..215R} and individual frequencies with a correction for the surface term by \citet{2014A&A...568A.123B}. The results from these approaches are in full agreement. 

Information on the stellar distance is incorporated \citep{2018MNRAS.475.5487S} to constrain the stellar modeling. We use the \textit{Gaia} DR2 parallax combined with $JHK_s$ photometry from the Two Micron All Sky Survey \citep[2MASS;][]{2003yCat.2246....0C,2006AJ....131.1163S} and extinction from the \citet{Green2019} dust map to determine absolute magnitudes, which are then fitted to the grid.


\tref{tab:model} gives the results from the BASTA run using ratios and incorporating information on the distance. We have added in quadrature to the uncertainties the difference between using ratios and individual frequencies in the modeling.

As a sanity check of the modeling, a grid-based model was computed using the Yale-Birmingham code \citep[YB; ][]{2010ApJ...710.1596B, 2012ApJ...746...76B,2011ApJ...730...63G}, which takes a different approach than BASTA and uses different grids of stellar models -- see \citet{keystone} for further details. The results from this approach agree fully within uncertainties with the BASTA results. 

As a further check we also compare the distance from \textit{Gaia} with that obtained by combining the stellar radius with the angular diameter from the IRFM (\tref{tab:spec}). For the BASTA results a distance of $106.0 \pm 1.9$ pc is obtained,
in agreement with the \citet{BJ2018} \textit{Gaia} distance of $d=106.29_{-0.67}^{+0.68}$ pc.

To estimate the size of a potential systematic uncertainty from different approaches and input physics in the modeling, besides the check using the YB code, we used the results from the \kp LEGACY study \citep{legacy,2017ApJ...835..173S} and determined the median of the scatter in central parameter values for stars similar to HIP 41378 (masses from $\rm 1.05$ to $1.35 M_{\odot}$ and radii from $\rm 1.15$ to $1.45\, R_{\odot}$). We find median relative systematic differences of $1.8\%$ in mass, $0.5\%$ in radius, $0.3\%$ in density, $0.06\%$ in $\logg$, and $4\%$ in age. These systematic differences are all below our quoted statistical uncertainties from BASTA of $2.5\%$ in mass, $0.7\%$ in radius, $1.0\%$ in density, $0.1\%$ in $\logg$, and $17\%$ in age. Given the minor contribution such systematic uncertainties would have on our reported estimates if joined with our statistical uncertainties and the uncertainty in their estimation (based here only on 17 similar stars from \kp), we do not include a systematic term on our quoted parameters nor in our further analysis.

\begin{deluxetable*}{lcccccccc}
\tablecaption{Results from the asteroseismic modeling.\label{tab:model}}
\tablehead{\colhead{Method} & \colhead{Mass} & \colhead{Radius} & \colhead{Density} & \colhead{\logg} & \colhead{Age} & \colhead{Distance} & \colhead{\teff} & \colhead{\feh} \\ 
 \colhead{} & \colhead{$\rm (M_{\odot})$} & \colhead{$\rm (R_{\odot})$} & ($\rm g/cm^3$) & \colhead{(cgs; dex)}  & \colhead{(Gyr)}  & \colhead{(pc)}  & \colhead{(K)}  & \colhead{(dex)} } 
\startdata
BASTA & $1.22^{+0.03}_{-0.02}$ & $1.300\pm0.009$ & $0.785\pm0.008$ &$4.298\pm0.004$ & $2.07^{+0.36}_{-0.27}$ & $106.8\pm1.0$ & $6290\pm77$ & $-0.05\pm 0.10$\\
\citetalias{K293}  & $1.15\pm 0.064$ & $1.4\pm 0.19$ &  -- & $4.18\pm 0.1$ & -- & $116\pm 18$ & $6199\pm 50$ & $-0.11\pm 0.08$ \\
\enddata
\end{deluxetable*}
\subsection{Planetary analysis}\label{sec:planet}
The periods of the two innermost planets (``b'' and ``c'') were already well-determined by \citetalias{K293} from C5 data. For the short period planet ``b'' we again detect multiple transits in C18, whereas planet ``c'' only transits once (see \fref{fig:ps}). Planets ``d'' and ``f'' both show a single transit in C18, while ``e'' does not transit during C18 \citep{bernado, becker}. Using the asteroseismic stellar parameters derived in this study, we can further improve on the properties of the planets in the system.

\subsubsection{Transit fitting}\label{sec:fit}
For fitting the transits we used the \citet{Mandel} model, calculated using the \texttt{BATMAN} package \citep{batman}. For the optimization of transit parameters this was combined with the Affine Invariant Markov Chain Monte Carlo sampler \texttt{EMCEE} \citep{emcee}. \texttt{BATMAN} was used adopting a quadratic limb-darkening law with Gaussian priors for the limb-darkening coefficients using the values from \citetalias{K293} with a width of $0.1$. 

\begin{deluxetable*}{lcccccccc}
\tablecaption{Planetary Parameters from Joint Transit Fit. \label{tab:planet}}
\tablehead{\colhead{Planet} & \colhead{$P$} & \colhead{$R_\mathrm{p}$} & \colhead{$a$} & \colhead{$a$} & \colhead{$i$} & \colhead{$b$} & \colhead{$t_\mathrm{d}$} &\colhead{$T_0$}  \\ 
 \colhead{} & \colhead{(days)} & \colhead{$(R_\star)$} & \colhead{$(R_\star)$} & \colhead{(AU)} & \colhead{(deg)} & \colhead{} & \colhead{(hours)} & \colhead{(BJD-2454833)}   } 
\startdata
``b'' & $15.57209\pm 0.00002$ & $0.0180^{+0.0002}_{-0.0003}$ & $22.8^{+1.3}_{-1.0}$ & $0.138^{+0.008}_{-0.006}$ & $89.2^{+0.7}_{-0.3}$ & $0.31^{+0.18}_{-0.22}$ & $5.06 \pm 0.03$ & $2319.283^{+0.001}_{-0.002}$  \\
``c'' & $31.7061^{+0.0001}_{-0.0002}$ & $0.0182\pm 0.0008$ & $36^{+6}_{-9}$ & $0.22^{+0.03}_{-0.06}$ & $88.6\pm 0.4$ & $0.90^{+0.06}_{-0.03}$ & $3.21 \pm 0.19$ & $2330.162\pm 0.003$  \\
``d'' & $278.360\pm 0.001$\textdagger & $0.0260^{+0.0004}_{-0.0006}$ & $190\pm 20$ & $1.17^{+0.14}_{-0.11}$ & $89.8\pm 0.1$ & $0.58^{+0.14}_{-0.09}$ & $12.44^{+0.10}_{-0.16}$ & $2333.273\pm 0.004$ \\ 
``e'' & $260^{+160}_{-60}$\textdagger & $0.037\pm 0.001$ & $112^{+14}_{-13}$ & $0.68^{+0.09}_{-0.08}$ & $89.7\pm 0.1$ & $0.52^{+0.19}_{-0.15}$ & $13.00^{+0.12}_{-0.15}$ & $2309.020\pm 0.001$  \\
``f'' & $542.0793\pm 0.0002$\textdagger & $0.0664\pm 0.0001$ & $230.6^{+1.3}_{-1.1}$ & $1.394^{+0.013}_{-0.012}$ & $89.96\pm  0.01$ & $0.17 \pm 0.03$ & $18.906^{+0.015}_{-0.016}$ & $2353.9162\pm 0.0003$  \\
\enddata
\tablecomments{\footnotesize For the limb-darkening coefficients we find $c_1 = 0.410^{+0.013}_{-0.014}$ and $c_2 = 0.12 \pm 0.02$. Value of $a$ for planet ``c'' is constrained from the period and assuming a zero eccentricity orbit. \newline \textdagger: Value constrained from \eqref{eq:pp} (see \fref{fig:pei}). We note that the uncertainty on the period for planets ``d'' and ``f'' does not reflect the width of the distributions in \fref{fig:pei}, because the period should correspond to one of the discrete periods given by \eqref{eq:pern}. For planet ``d'', a value of $n=4\pm 1$ (\eqref{eq:pern}) better represents the uncertainty in the predicted period. For ``e'' we estimate the period from the distribution in \fref{fig:pei}.}
 \end{deluxetable*}

The orbital parameters, \ie, period, semi-major axis ($a/R_{\star}$), mid-transit time ($T_0$), radius ratio ($R_p/R_{\star}$), and inclination, were fitted using uninformative flat priors. The starting point for the adopted 100 walkers were values close to those found in \citetalias{K293}, except for the periods of planets ``d'', ``e'', and ``f'' as described below. To account for the K2 cadence and the difference in cadence between the data used from the two campaigns (LC in C5 and SC in C18), the model light curves were oversampled by factors of 10 (SC) and 300 (LC) and then binned to the cadence of the observations.

In our fitting we assumed an eccentricity of zero, but discuss in \sref{sec:ecc} possible constraints on the eccentricity. We adopted a zero eccentricity, because the asymmetry from an eccentric orbit would be too small to properly constrain from the K2 photometry, as also noted in \citetalias{K293}. Following \citet{2010arXiv1001.2010W} the difference in ingress ($\tau_{\rm ing}$) and egress ($\tau_{\rm egr}$) time, causing the transit to appear asymmetric, can to leading order in $R_{\star}/a$ and $e$ be given as
\begin{equation}
   \mathcal{A}\equiv \frac{\tau_{\rm egr} - \tau_{\rm ing}}{\tau_{\rm egr} + \tau_{\rm ing}} \sim e\cos \omega \left( \frac{R_{\star}}{a} \right)^3 (1-b^2)^{3/2}\, .
\end{equation}
As an example, the innermost planet ``b'' of the system with an $R_{\star}/a\approx 0.04$ will have $\mathcal{A}<1\times 10^{-4}\, e$. For planet ``f'', with $R_{\star}/a\approx 0.0043$ (assuming the period found in \sref{sec:period}) the value for the asymmetry will be $\mathcal{A}<8.3\times 10^{-8}\, e$. Additionally, from our assessment in \sref{sec:ecc} of the constraints that can be put on $e$ from having the asteroseismic value for the stellar density, we find that the argument of periastron ($\omega$) in the eccentric cases would be close to ${\sim}270^{\circ}$. In this case $\cos \omega$ would tend to zero, further decreasing the asymmetry of the transit. 

Initially, each planet was fitted independently. For each iteration of the fitting we added a step to eliminate possible residual systematics from the light curve detrending, by fitting a linear slope in addition to the model light curve for each transit for a given planet. For the initial fits we ran the sampler for $10,000$ steps with a burn-in of $5000$ steps. 

For planets ``d'' and ``f'' there are several allowed periods (see \eqref{eq:pern}). We fitted the transits assuming each of these allowed periods to test the impact on other transit parameters. To prevent a walker from jumping to an allowed period other than the one being tested, we constrained the period to a small interval around the tested value. We further adopted a parallel tempering approach in the MCMC, with 10 different temperatures for each of the walkers. 

\begin{figure*}
\noindent
\centering
\makebox[\textwidth]{\includegraphics[width=1.1\textwidth]{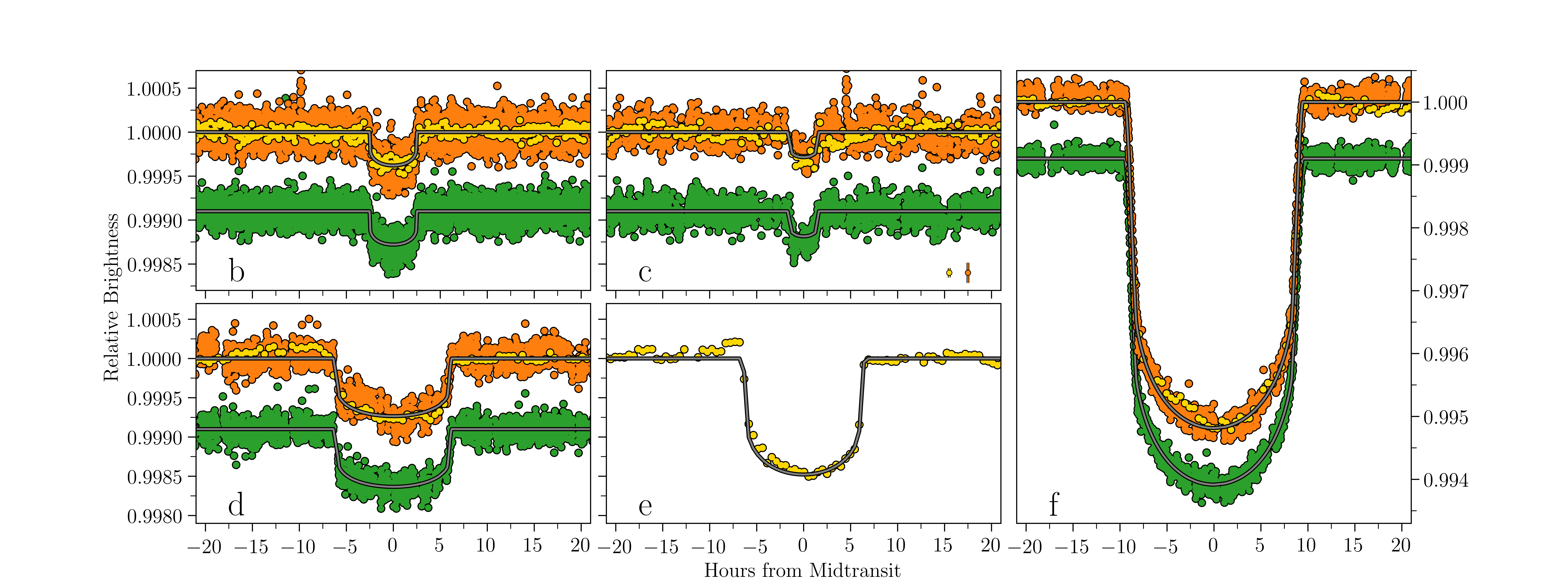}}
\caption{Phased light curves for all five planets using the parameters in \tref{tab:planet}, and with the fitted model overplotted. Data from \textit{K2P$^2$} are displayed in yellow, with C5 LC data given in light colors and C18 SC in dark. The data shown in green, and offset vertically from the yellow points, shows the C18 SC data from the \textit{K2SFF} pipeline \citep{vanderburg2014} -- the fitted model (applied to the yellow points) has been offset by the same amount for a more direct comparison. Planets ``b'' and ``c'' are displayed in the top panels and ``d'' and ``e'' are in the bottom panels. Shown to the right is planet ``f''. Denoted in the panel for planet ``c'' are representative photometric uncertainties for the LC and SC data. }
\label{fig:lcs}
\end{figure*}

A final joint fit including all planets was run after having constrained the starting values from the individual fits. The convergence and mixing of the walkers for this final run was assessed by visual inspection, and making use of the Gelman-Rubin convergence diagnostic \citep{gelman1992} and checking the effective sample size \citep{geyer1992}. For this we used the routines available in \texttt{PyMC3} \citep{pymc3}. Final planetary parameters are given in \tref{tab:planet}. \fref{fig:lcs} shows a phase plot for the planets together with the fitted transit light curve. In each panel the signal from the other planets have been removed. 

A study by \citet{grunblatt2016} showed that planet parameters modeled from K$2$ light curves can widely vary depending on the pipeline used to reduce the data. Therefore, we checked the consistency of our derived parameters by fitting for the planet properties on a light curve produced with \textit{K2SFF} \citep{vanderburg2014} with the systematics correction fit rederived by fitting simultaneously with the transits \citep{2016ApJS..222...14V}.
The \textit{K2SFF} short-cadence light curve is shown in Figure \ref{fig:lcs} in green, with a slight offset from the \textit{K2P$^2$} light curves for a direct, visual comparison. 

First described by \citet{chontos2019}, this independent analysis fitted for the following parameters: orbital period ($P$), time of mid-transit ($T_0$), linear ($c_1$) and quadratic ($c_2$) limb-darkening coefficients, mean stellar density assuming a circular orbit ($\rm \rho_{\star, circ}$), impact parameter ($b$), ratio of the planetary radius to the stellar radius ($R_p/R_{\star}$), and the photometric zero-point ($z$). To keep the two analyses consistent, the same priors are imposed and the MCMC samplers are run with the same amount of walkers and steps, including the same burn-in. When using light curves reduced through two different pipelines and modeled through two independent analyses, the parameters still agree to within 1$\sigma$ for all derived quantities and thus provides further evidence for the validity of the derived planet properties.
\subsubsection{Planetary periods}\label{sec:period}
\begin{figure*}
\centering
\includegraphics[width=1\textwidth]{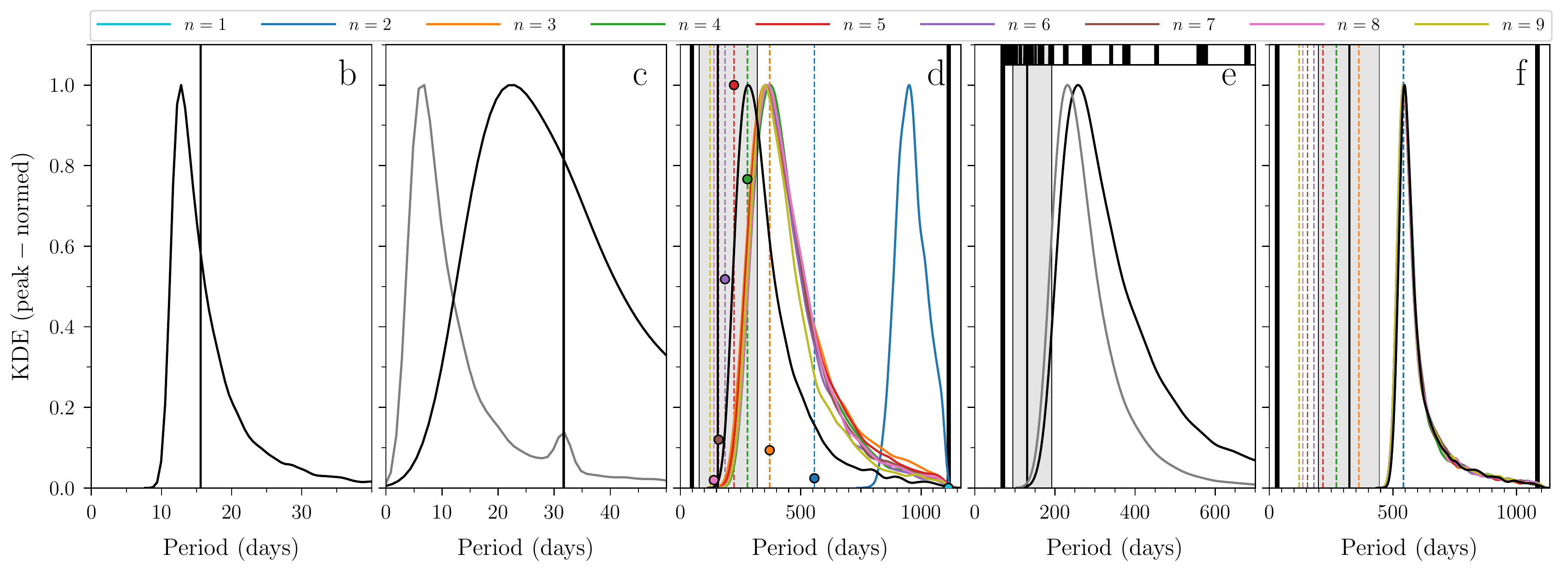}
\caption{Period distributions resulting from solving \eqref{eq:pp} for all five planets going from ``b'' in the leftmost panel to ``f'' in the rightmost. The vertical black lines in the gray shaded areas mark the values from \citetalias{K293}. The thick black lines denote the minimum and maximum allowed value for the periods of ``d'', ``e'', and ``f''. The black curves have been made using the parameters from the joint fit in \tref{tab:planet}. The gray curve for ``c'' has been created using parameters from a fit without constraints on $\frac{a}{R_\star}$. The color coding for ``d'' and ``f'' refers to the assumed period when performing the individual fits to obtain values for the parameters entering in \eqref{eq:pp} and the dashed vertical lines denote the period assuming a value for $n$. For ``d'' the markers indicate the relative posterior probabilities evaluated at the positions of the discrete periods after applying the normalized prior probabilities by \citet{becker} (their Table~2). For ``e'' the horizontal bar in the top shows the allowed periods in white, whereas the black areas are the forbidden periods as these would have resulted in a transit in C18 or in sector 7 of TESS (see \fref{fig:lctess}); the gray line shows the period distribution after applying a prior on the period probability following \citet[][their  Eq.~1]{becker}.}
\label{fig:pei}
\end{figure*}

As done by \citetalias{K293} we predict the planetary periods from information of the star and parameters from the transit fit. 
The planetary period can be obtained by solving the following relation:
\begin{align}\label{eq:pp}
    t_{\mathrm{d},i}=&\frac{P_i}{\pi}\arcsin \Biggl[ \left( \frac{G(M_{\star} + m_{\mathrm{p},i})P_i^2}{4\pi^2} \right)^{-1/3}  \\\notag
    &  \times \sqrt{(R_{\mathrm{p},i} + R_{\star})^2 - b_i^2R_{\star}^2} \Biggr]\,  \frac{\sqrt{1-e_i^2}}{1 + e_i \cos \omega_i}\, ,
\end{align}
where $t_\mathrm{d}$ is the transit duration, $m_\mathrm{p}$ is the mass of the planet, $b$ is the impact parameter, $\omega$ is the argument of periastron, $e$ is the eccentricity, and the subscript $i$ refers to a given planet. By drawing samples from a normal distribution created from each of the stellar and transit parameters and their errors, we can thus build a distribution for the period. The obtained period distributions are shown in \fref{fig:pei}. For the eccentricity we adopt a $\beta$-distribution \citep[with parameters $\alpha=0.867$ and $\beta=3.03$ from][]{2013MNRAS.434L..51K} and fixed $\omega$ to $3\pi/2$ (see \sref{sec:ecc}).
The masses in \eqref{eq:pp} are estimated using the mass-radius relation in \citet{Wolfgang2016} with two different power laws, one for planets with $R_\mathrm{p} < 4 R_\Earth$ and another for planets in the range  $4 R_\Earth < R_\mathrm{p} < 8 R_\Earth$. For planets larger than $8 R_\Earth$ we uniformly draw samples from a Jovian density distribution, \ie, $\rho = 1.3 \pm 0.5$ g cm$^{-3}$ consistent with the radius anomaly reported in \citet{Laughlin2011} which should not be relevant for the present case. 

For the two inner planets for which the period is well-established, the period prediction serve as a sanity check of the fitted transit parameters. For planet ``b'' the measured and predicted period are seen to be in good agreement. 
For planet ``c'' we obtain transit parameters from the K2 data that result in a period distribution which poorly matches the measured period. We attribute this inconsistency to the quality of data at the transit times for planet ``c'', leading to a rather uncertain estimation of $a/R_\star$. We note that adopting the $a/R_\star$ and stellar parameters from \citetalias{K293} leads to a similar distribution. Assuming an orbit with zero eccentricity the $a/R_\star$ should be ${\sim}39$ rather than the value of $73$ reported in \citetalias{K293}. We therefore confined $a/R_\star$ to be in the interval $[25,\, 53]$ for planet ``c'' in the final fit. This added constraint leads to a predicted period in better agreement with the measured value (see \fref{fig:pei}).

For planets ``d'' and ``f'' we can constrain the periods to be the difference between the mid-transit times in C5 and C18 divided by an integer, \ie,
\begin{equation}
    P_n = \frac{T_{0,\mathrm{C18}}-T_{0,\mathrm{C5}}}{n} \, , \quad n=1,2,3,\dots \, ,
    \label{eq:pern}
\end{equation}
where $T_{0,\mathrm{C18}}$ and $T_{0,\mathrm{C5}}$ are the mid-transit times observed in C18 and C5. Given the $\sim 3$ yr gap between C5 and C18 this gives some 20 possible periods for each, with a lower boundary from the lack of additional transits in the individual time series. 

The importance of the precision of the stellar parameters for the resulting distributions for the period and eccentricity differs from planet to planet. The impact is, for instance, much more pronounced for planet ``f'', where the parameters $t_\mathrm{d}$, $b$, and $R_\mathrm{p}$ can be determined with great precision (see \tref{tab:planet}).

We see the that the predicted period from the individual planet fits is stable against the use of transit-fit parameters based on different input periods. With the exception of one, the resulting periods for planet ``d'' all end up at a period corresponding to a value of $n=3$ ($P\simeq371$ days), which suggests two missed transits between C5 and C18. The same is true for planet ``f'', where the periods match the allowed period corresponding to $n=2$ ($P\simeq542$ days), indicating that a single transit has been missed between C5 and C18. From the final joint fit, the parameters for planet ``d'' results in a period that is lower than from the individual fits, corresponding to a value of $n=4$ ($P\simeq278$ days). The change in period is attributed to a change in transit parameters from the better constraint on limb-darkening parameters in the joint fit, where we find $c_1 = 0.410^{+0.013}_{-0.014}$ and $c_2 = 0.12 \pm 0.02$ for the linear and quadratic limb-darkening coefficients, respectively. We note that these coefficients are in good agreement with table values from \citet{2018yCat..36180020C}. This is especially true for the linear limb-darkening coefficient. While $n=4$ provides the best match for planet ``d'' the predicted period distribution is relatively broad, meaning that periods of $223$ days ($n=5$) and $371$ days ($n=3$) cannot be excluded. The periods we report in \tref{tab:planet} for planet ``d'' and ``f'' are thus to be taken as given this value for $n$ (corresponding to an orbit with low eccentricity) what would then be the precision with which the period can be determined. A value of $n = 4 \pm 1$ better represents the uncertainty for the period of planet ``d''.

We note that the $n=4$ period for ``d'' and $n=2$ period for ``f'' both correspond to low-eccentricity orbits (see \fref{fig:ecc} and \sref{sec:ecc}), consistent with the circular orbits adopted in the transit fit.
We did also try adopting eccentricity distributions derived in \sref{sec:ecc} for the period distributions -- this causes small shifts in the individual distributions, leaving only the low-eccentricity solutions with predicted periods consistent with the input period. 

The periods corresponding to $n=2$ for ``f'' and $n=4$ for ``d'' that we use for our final fit are not the most likely periods according to \citet{becker}, however, this configuration is found to be dynamically stable in both \citet{becker} and \citet{bernado} and as such are consistent with their results. The same goes for having $n=3$ or $n=5$ in \eqref{eq:pern} for ``d''.

Stability calculations for configurations similar to the one we report in \tref{tab:planet} have been carried out by \citet{becker} and \citet{bernado}, but as both have used different stellar parameters we used our parameters and tested the stability of the orbital solution using the \texttt{Mercury6} $N$-body integrator \citep{mercury}. We found that this solution was stable for at least the integrated 100 Myr years.

\begin{figure}
\includegraphics[width=\columnwidth]{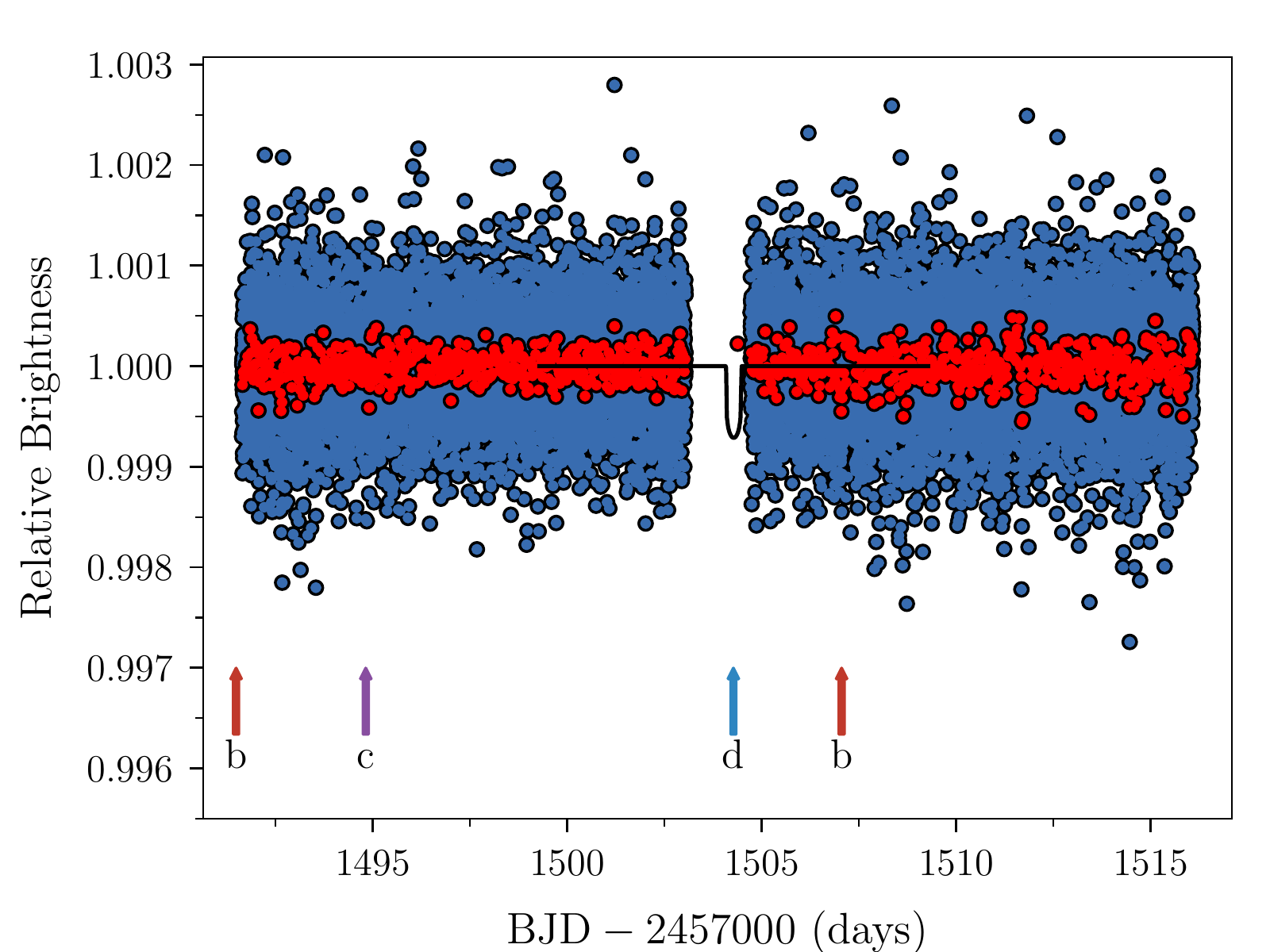}
\caption{Light curve for HIP 41378 obtained during sector 7 of the TESS mission. The blue points show the 2 minute cadence photometry, having only applied a simple detrending to the data using the \texttt{Lightkurve} module. The red points show the data when binned to a cadence of 30 minutes. Using the ephemerides and possible periods from K2 the corresponding transit times for each planet have been indicated with arrows. For planet ``d'' the only possible period that would result in a transit during the sector 7 baseline is that of $n=5$. The black transit model shows transit shape for planet ``d'' based on the K2 analysis.}
\label{fig:lctess}
\end{figure} 

As mentioned by \citet{becker} the K2-93 system was to be observed by TESS. \fref{fig:lctess} shows the light curve, with a 2 minute cadence, extracted from TESS sector 7 data. The light-curve extraction and a simple detrending was done using the \texttt{Lightkurve} \citep{2018ascl.soft12013L} module. The scatter in the TESS data is much higher than for K2, and while difficult to see, planets ``b'' and ``c'' both transit once. In \fref{fig:lctess} we have also indicated the possible transit time for planet ``d'' according to \eqref{eq:pern} and the $T_0$ from the K2 data. Of the outer planets, planet ``d'' is the only one of the outer planets with a possible transit during the sector, with a period corresponding to $n=5$ (${\sim}223$ days). Unfortunately, the mid time for this transit falls within the TESS downlink gap. As also indicated by the black transit model, which is based on the transit fit to the K2 data, the entire transit is contained within the gap, \ie, one should not expect the see the egress in the TESS data. In sum, the TESS data cannot exclude nor confirm any of the possible periods for planets ``d'' and ``f''. An extra constraint on allowed periods can, however, be placed for planet ``e'', which we have included in \fref{fig:pei}.

Having only observed a single transit for planet ``e'' in both campaigns, we obtain a range of possible periods, as opposed to the discrete set for ``d'' and ``f''. Furthermore, these extend beyond the difference in time between the two campaigns. We have a lower limit for the period given as the difference between the mid-transit time and the end of C5. Also, periods that would result in a transit in C18 or in sector 7 of the NASA TESS mission \citep{2014SPIE.9143E..20R} can be excluded (see Figures~\ref{fig:pei} and \ref{fig:lctess}). In the transit fitting we adopted the predicted period from \eqref{eq:pp} of $\sim260$ days, because this is the allowed period from \eqref{eq:pp} consistent with a low-eccentricity orbit.

As noted by \citet{becker} not all of the possible periods governed by \eqref{eq:pern} are equally likely, because it is more likely that we would observe transits for ``d'' and ``f'' in C18 for larger values of $n$.
For planet ``d'' \citet{becker} provides normalized probabilities for the discrete periods, considering both the likelihood of the periods given the number of observed transits combined with the observation baseline and the dynamical stability of the orbits (their Table~2). If we apply these probabilities to the period distribution from our joint fit evaluated at the discrete periods we find that the posterior probability of the $n=5$ solution becomes slightly higher than $n=4$ where our distribution peaks (see \fref{fig:pei}). The result from this is thus still consistent with the conservative estimate on the period uncertainty of $n = 4 \pm 1$ for planet ``d''. We caution that the dynamical stability calculations of \citet{becker} used stellar parameters different from the ones provided by the asteroseismic analysis which could influence the prior period probabilities. For planet ``e'' we applied only the prior based on the baseline \citep[][their Eq.~1]{becker}, which moved the distribution peak slightly lower -- the resulting period of $230^{+120}_{-60}$ days is fully consistent with the quoted period in \tref{tab:planet}.
For planet ``f'' the period distribution from \eqref{eq:pp} is so well-constrained around $n=2$ that the application of prior probabilities has no effect on the favored period.

\subsubsection{Eccentricity}\label{sec:ecc}
While we have assumed $e=0$ in our transit fitting, based on arguments presented in \sref{sec:fit}, it is interesting to see which constraints can be put on the eccentricity from having an asteroseismic estimate of the stellar density.
\begin{figure*}
\centering
\includegraphics[width=1\textwidth]{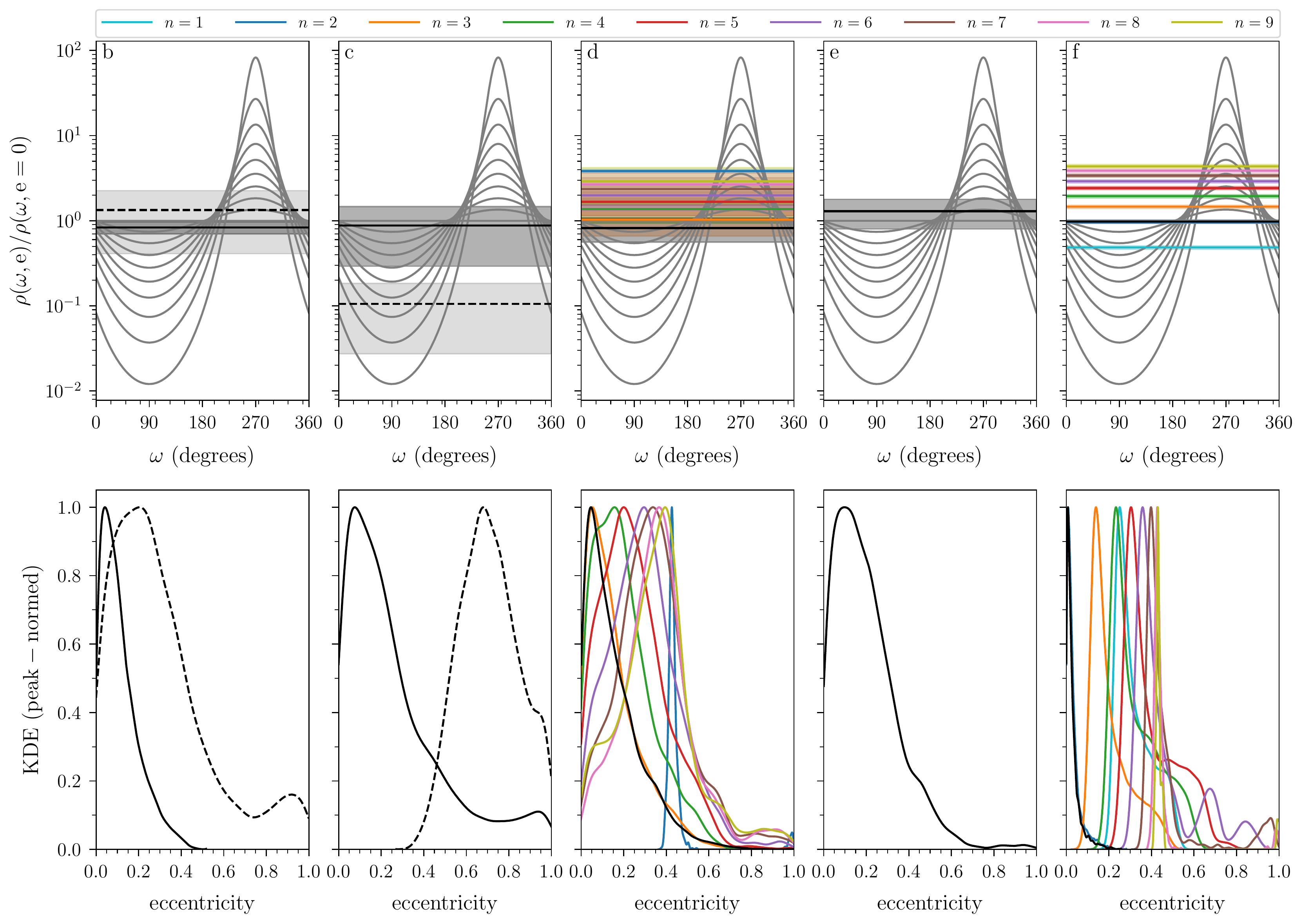}
\caption{Gray solid lines in the top panels display the curves governed by \eqref{eq:denrat}. The horizontal line and corresponding shaded area is our result for a planet in a given configuration. The resulting eccentricity distributions are shown in the panel below. The black solid lines show our results using parameters from the parameters in \tref{tab:planet} and the black dashed lines in ``b'' and ``c'' are the results obtained using the stellar parameters from \citetalias{K293}. As in \fref{fig:pei} the colored solid lines for ``d'' and ``f'' show the results from trying periods governed by \eqref{eq:pern} and only fitting for these particular planets. Note that results for smaller periods ($n > 9$) are not displayed, since the ratio in \eqref{eq:denrat} was significantly different from 1, making it difficult to obtain reasonable eccentricity distributions when marginalizing over $\omega$. The same is true for the $n=1$ case for planet ``d''.}
\label{fig:ecc}
\end{figure*} 

Assuming a circular orbit, the mean stellar density from a transiting planet, $\rho_{\star,\mathrm{transit}}$, can be estimated as \citep{Seager2003}:
\begin{equation}
    \rho_{\star,\mathrm{transit}} = \frac{3 \pi}{G P^2} \left ( \frac{a}{R_\star} \right)^3 \, ,
    \label{eq:rhotra}
\end{equation}
where $G$ is the gravitational constant. With an independent estimate of the stellar density, as the one obtained through the asteroseismic modeling, the ratio to the density in \eqref{eq:rhotra} can be written as \citep{Dawson2012,2014ApJ...782...14V,2015ApJ...808..126V}:
\begin{equation}
    \frac{\rho_{\star,\mathrm{astero.}}}{\rho_{\star,\mathrm{transit}}} = \frac{(1-e^2)^{3/2}}{(1+e \sin \omega)^3} \, ,
    \label{eq:denrat}
\end{equation}
that is, as a function of the orbital eccentricity $e$ and argument of periastron $\omega$. \fref{fig:ecc} shows for each planet the curves described by \eqref{eq:denrat} for different values of $e$. The values obtained for the density ratio are indicated by horizontal lines --- for the outer planets ``d'' and ``f'' each of the obtained ratios from different assumed periods (and thus different $a/R_\star$; \eqref{eq:pern}) are given. We do not show all ratios for ``d'' and ``f'', because the ratio in \eqref{eq:denrat} for periods with $n>9$ all correspond to highly eccentric orbits $(e>0.5)$. This is also true for the case of a single missed transit for planet ``d'', \ie, $n=1$. Furthermore, as reported by \citet{becker} periods for ``f'' corresponding to $n>6$ are very unlikely as they should have been detected by their ground-based observations.

Based on the density ratio we can compute the resulting probability density for the eccentricity by marginalizing over the possible values for $\omega$. This is done by a Monte Carlo sampling in $\omega$, where for each draw we also draw from the density ratio and then solve for the eccentricity that matches this value --- these distributions are shown in the bottom panels of \fref{fig:ecc}. In addition to a distribution we also obtain a range of possible $\omega$ values.  

Note that we cannot readily obtain a distribution for the eccentricity of planet ``e'' because the information on the period of the planet is not precise enough. The distribution shown in \fref{fig:ecc} is therefore obtained from assuming the period distribution from \eqref{eq:pp} (see \fref{fig:pei}). 

Our initial fit for planet ``c'' with no constraints on $a/R_\star$ resulted in a distribution, which was centered at a very high eccentricity for a close-in orbit ($P\simeq 31.7$ days) in a near $2:1$ resonance with planet ``b''. Assuming that the eccentricity should be close to $0$, as seen for planet ``b'', indicates that the $\frac{a}{R_\star}$ is underestimated. We note also, that if we adopt the value for $\frac{a}{R_\star}$ from \citetalias{K293} the resulting value for the eccentricity of planet ``c'' is even higher, with a value of $e\simeq 0.7$ as seen in \fref{fig:ecc}. As also mentioned in \sref{sec:period} we attribute this likely underestimation of $\frac{a}{R_\star}$ to a poor quality of the K2 data for the three transits obtained for this planet. \citet{bernado} observed transits of planets ``b'' and ``c'' with the {\it Spitzer Space Telescope} \citep[][]{spitzer} between C5 and C18. While their planetary parameters from K2 and {\it Spitzer} for ``b'' are well-determined and in agreement (also with the values we report in \tref{tab:planet}), $a/R_\star$ and $i$ for planet ``c'' are, as is the case for our fit, rather poorly constrained.

\begin{figure*}
\centering
\includegraphics[width=1.0\textwidth]{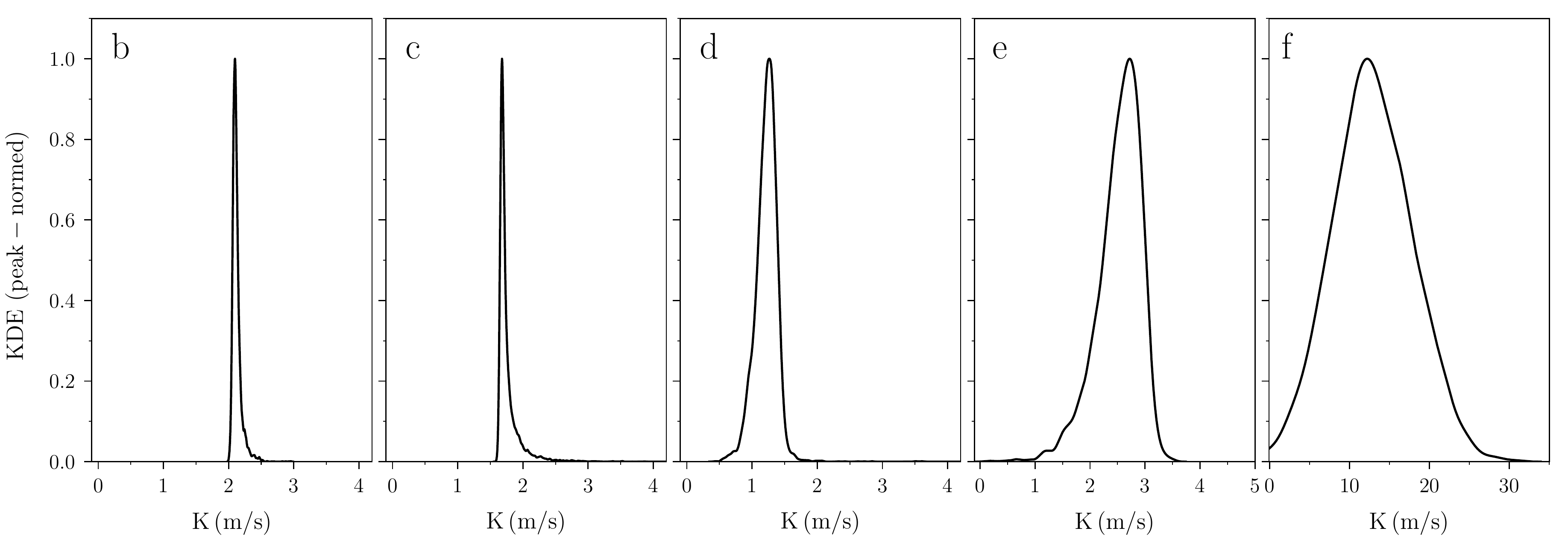}
\caption{Distributions for the RV amplitude estimated from \eqref{eq:kamp} for each planet again going from ``b'' on the left to ``f'' on the right. All curves have been created using the period distributions in \fref{fig:pei} resulting from the parameters in \tref{tab:planet}.}
\label{fig:kamp}
\end{figure*} 
Since the eccentricity cannot be properly constrained from the photometry alone, we cannot draw any firm conclusions on the possibility of non-zero eccentric orbits in the K2-93 system. We note, however, that when assuming a low eccentricity, periods consistent with the allowed periods can be obtained for planets ``d'' and ``f''. While the likelihood of eccentric solutions could be appraised from stability calculations, the sheer number of possible parameter combinations in a five-planet system, including a sampling of starting orbital phases, makes the problem somewhat intractable -- and, in any case, beyond the scope of this paper. Proper constraints on the eccentricity should rather be obtained from radial velocity follow-up, which would also place better constraints in the planetary masses. 

\section{Discussion}

The K2-93 system is very interesting for radial velocity (RV) follow-up, both because it is bright but also because systems with 5 or more planets are rare. With new stellar and planetary parameters we can update the estimate by \citetalias{K293} for the radial velocity amplitude of the star, $K$, induced by the planets. This amplitude can be estimated using the following relation:
\begin{equation}
    K = \frac{1}{(1-e^2)^{1/2}} \left ( \frac{2 \pi G}{P} \right )^{1/3} \frac{m_\mathrm{p} \sin i}{(M_\star + m_\mathrm{p})^{2/3}} \, ,
    \label{eq:kamp}
\end{equation}
which we do using Monte Carlo sampling. For planets ``b'' and ``c'', where we have well-determined periods, we sample from a normal distribution using the periods in \tref{tab:planet}. For ``d'', ``e'' and ``f'' we sample from the distributions obtained from the final fit in \fref{fig:pei}. For the eccentricities we draw from the distributions in \fref{fig:ecc} (again, respectively, using $n=4$ and $n=2$ for ``d'' and ``f''), except for planet ``c'', where we draw from a $\beta$-distribution due to the complications mentioned in \sref{sec:period}. The resulting distributions for the $K$-amplitude induced by each planet is seen in \fref{fig:kamp}. The planetary masses are obtained from the mass-radius relation in \citet{Wolfgang2016}.
For planets ``b'' and ``c'' our estimate of the RV semi-amplitude is similar to that of \citetalias{K293}, with values of the order $1.5-2.2\, \, \rm m/s$. For the outer planets we predict lower amplitudes, because our predicted periods are in all cases larger than those estimated by \citetalias{K293}. Even though the semi-amplitudes are lower the tighter constraints on the periods should aid in the isolation of the RV contributions from the individual planets. The rather small semi-amplitudes for the planets ``b'' through ``e'' is still achievable with the current generation of high-precision radial velocity spectrographs.

\begin{figure}
\centering
\includegraphics[angle=-90,width=0.95\columnwidth]{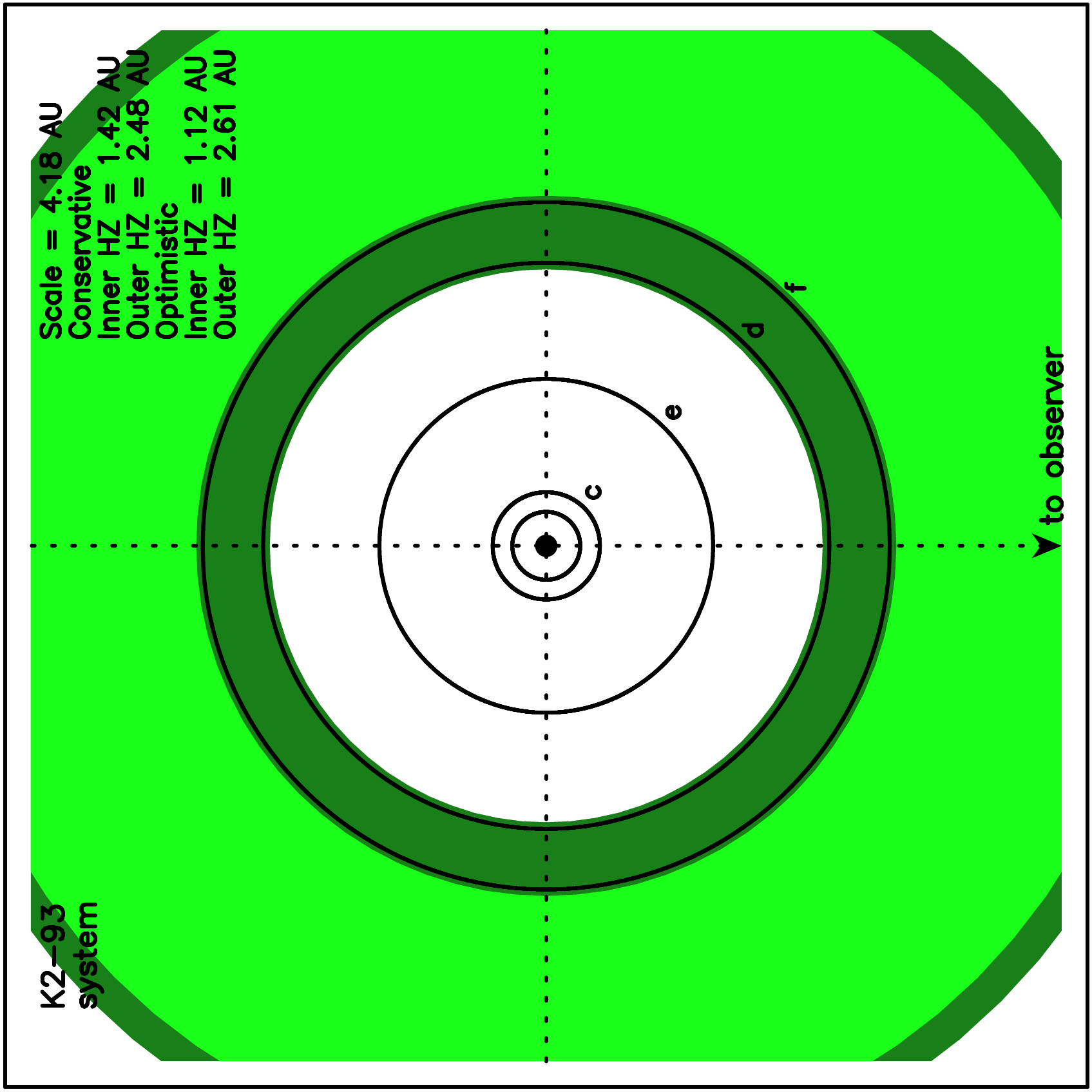}
\caption{A top-down view of the K2-93 planetary system spanning $4.18$ AU across, where the orbits of the planets are shown as solid circles. The conservative HZ is shown in light green, and the optimistic extension to the HZ is shown below in dark green.}
\label{fig:hz}
\end{figure} 
Using the improved stellar properties from \tref{tab:model} and derived planetary properties from \tref{tab:planet}, we calculate the extent of the Habitable Zone (HZ) and place the planetary orbits within that context, see \fref{fig:hz}. Such calculations are important given the combination of revised stellar properties and improved stellar distances \citep{kane2018}. To calculate the HZ boundaries, we use the formalism described by \citet{kopparapu2013} and \citet{kopparapu2014}. These boundaries include optimistic (using assumptions regarding past surface liquid water on Venus and Mars) and conservative (runaway greenhouse and maximum greenhouse) scenarios \citep{kasting2014,kane2014,kane2016}. We calculate the boundaries for the K2-93 inner optimistic HZ, inner conservative HZ, outer conservative HZ, and outer optimistic HZ as $1.12$, $1.42$, $2.48$, and $2.61$ AU respectively. Most of the planets lie interior to the inner optimistic HZ boundary, the so-called ``Venus Zone'' dominated by runaway greenhouse atmospheres for terrestrial planets \citep{kane2014}. The outermost planets (``d'' and ``f'') lie within the inner part of the optimistic HZ based on these revised calculations. During the evolution of the K2-93 host the HZ boundaries have shifted compared to when the star was on the zero-age MS (ZAMS). The increase in luminosity and decrease in \teff has the net effect of gradually moving the HZ boundaries outward, particularly when the star moves off the MS \citep{gallet2017}. Since the outermost planets lie within the inner regions of the HZ, they are likely to have occupied the conservative HZ early in the MS lifetime of the star. The planets range in size from mini-Neptune to Jupiter (planet ``f'') and thus are likely giant planets. However, giant planets within the HZ are interesting from the perspective of potential exomoon habitability \citep{hinkel2013,heller2014} and the occurrence rates of HZ giant planets hava been shown to be relatively low \citep{hill2018}. The detection of exomoons is a difficult endeavor and has been undertaken using transit signatures in the precision photometry from
the \kp mission \citep{kipping2009,kipping2012}. However, such exomoon searches lie at the threshold of detectability and can
lead to ambiguous interpretations of the data \citep{teachey2018,kreidberg2019}. \citet{hill2018} provide estimates for the expected angular separation of exomoons from their host planets, the detection requirements of which are beyond the capabilities of present facilities. Significant consideration has been
applied to the theory and methodology of biosignature detection for terrestrial planets \citep{fujii2018,schwieterman2018}, but the techniques of transmission spectroscopy and direct imaging will be likewise inhibited by low signal-to-noise observations in the near future.

The question arises as to whether the presence of the giant ``f'' planet located at the inner boundary of the optimistic and conservative HZ regions excludes stable orbits for HZ terrestrial planets in the system. To estimate this, we calculate the mutual Hill radii for adjacent planet pairs:
\begin{equation}
  R_{H,M_p} = \left[ \frac{M_{p,in} + M_{p,out}}{3 M_\star}
  \right]^{\frac{1}{3}} \frac{(a_{in} + a_{out})}{2}
    \label{mhillradius}
\end{equation}
where ``in'' and ``out'' refer to the inner and outer planets in an adjacent pair \citep{crossfield2015,sinukoff2016}. Using the stability
criterion of $\Delta = (a_{out} - a_{in})/R_H > 9$ for adjacent planets \citep{smith2009}, we estimated the smallest semi-major axis
for an Earth-mass planet exterior to planet ``f'' that can fulfill the criterion, assuming a Jupiter mass for planet ``f''. Our calculations
show that this minimum semi-major axis is located at ${\sim}2.52$~AU, placing such a hypothetical planet in the outer part of the optimistic
HZ region (see \fref{fig:hz}). Thus it is still (barely) possible for a terrestrial planet to retain orbital integrity within the HZ of the
system.


\section{Conclusions}
We have re-analyzed the K2-93 multi-planet system, which was discovered and first analyzed by \citetalias{K293} based on long-cadence data from K2 Campaign 5. Short-cadence data obtained during K2 C18 have enabled us to perform an asteroseismic analysis of the host star, placing strong constraints on the stellar parameters. From the asteroseismic modeling we obtain a value for the stellar mass of $1.22^{+0.03}_{-0.02}\, \rm M_{\odot}$, a stellar radius of $1.30\pm 0.01\, \rm R_{\odot}$, and an age of $2.07^{+0.36}_{-0.27}$ Gyr. The asteroseismic analysis further suggests that a high obliquity can be ruled out, but the stellar inclination can only be weakly constrained.

The updated stellar parameters from our asteroseismic analysis have enabled an improved prediction of the periods of planets ``d'' and ``f'', which for both planets match one of their allowed periods. We predict the period of planet ``d'' to be ${\sim}278$ days, but note that the value of $n$ in equation \eqref{eq:pern} has an uncertainty of $\pm 1$. For planet ``f'' we predict the period to be ${\sim}542$ days, while for planet ``e'' (which did not transit again in C18) we predict a period of ${\sim}260^{+160}_{-60}$ days.
To appraise the impact on the planetary analysis from the adopted data reduction, we applied an independent analysis \citep{chontos2019} to the C18 data obtained from the \textit{K2SFF} pipeline \citep{vanderburg2014}. This independent analysis returned parameters in agreement with the results reported in \tref{tab:planet}.

We found that the transit parameters of planet ``c'' are poorly determined using the K2 data, resulting in an unrealistic value for the orbital eccentricity and a predicted period that does not match the measured period. Comparing with \citetalias{K293}, we find that their reported transit parameters appear to suffer the same problem. It would therefore be interesting to observe the transit of planet ``c'' from a ground-based facility, which should be possible given the period of only ${\sim}31.7$ days though the shallow transit for this planet would make it difficult to detect. 

By comparing the stellar density determined from asteroseismology with that obtained from the transit fit we computed distributions for the orbital eccentricities of the planets. For all planets we predict low-eccentricity orbits, and find in particular for planets ``d'' and ``f'' that the low-eccentricity solution results in a predicted period consistent with an allowed period. 
A better constraint on the orbital eccentricities, and the planetary masses, should be obtained from radial velocity observations. 

Based on our updated stellar and planetary parameters we found that planets ``d'' and ``f'' fall within the inner part of the optimistic habitable zone, making these planets interesting in terms of potential exomoon habitability.

\section*{Acknowledgements}
We thank the anonymous referee for comments and suggestions that helped improve the paper.
The authors acknowledge the dedicated teams behind the \kp and K$2$ missions, without whom this work would not have been possible. Funding for the K2 mission is provided by the NASA Science Mission directorate. Short-cadence data were obtained through the Cycle-$6$ K$2$ Guest observer program. 
Funding for the Stellar Astrophysics Centre is provided by The Danish National Research Foundation (Grant agreement no.: DNRF$106$).
MNL acknowledges the support of ESA PRODEX programme.
VSA acknowledges support from VILLUM FONDEN (research grant 10118)  and the Independent Research Fund Denmark (Research grant $7027$-$00096$B).
SB acknowledges support from NASA grant $80$NSSC$19$K$0102$ and NSF grant AST-1514676.
CvE acknowledges support from the European Social Fund via the Lithuanian Science Council grant No. $09.3.3$-LMT-K-$712-01-0103$. 
WJC and GRD acknowledge the support of the UK Science and Technology Facilities Council (STFC).
LC is the recipient of the ARC Future Fellowship FT160100402.
This work has made use of data from the European Space Agency (ESA) mission Gaia (\url{https://www.cosmos.esa.int/gaia}), processed by the Gaia Data Processing and Analysis Consortium (DPAC, \url{https://www.cosmos.esa.int/web/gaia/dpac/consortium}). Funding for the DPAC has been provided by national institutions, particularly the institutions participating in the Gaia Multilateral Agreement.

\facilities{\kp/K2, Gaia DR2, \href{https://archive.stsci.edu/index.html}{MAST}, \href{http://kasoc.phys.au.dk/index.php}{KASOC}, \href{https://tasoc.dk/}{TASOC}, TESS}

\software{astropy \citep{Astropy},  \texttt{Lightkurve} \citep{2018ascl.soft12013L}, \texttt{Emcee} \citep{emcee}, \texttt{BATMAN} \citep{batman}, \texttt{PyMC3} \citep{pymc3} }

\vspace{1cm}
\small
\bibliography{MASTERBIB}
\label{lastpage}

\end{document}